\documentclass[sigplan,screen]{acmart}

\usepackage{dirtytalk}
\usepackage{hyperref}
\usepackage{cprotect}
\usepackage{listings}
\usepackage{fancyvrb}

\AtBeginDocument{%
  \providecommand\BibTeX{{%
    \normalfont B\kern-0.5em{\scshape i\kern-0.25em b}\kern-0.8em\TeX}}}

\setcopyright{none}
\acmYear{2023}
\acmDOI{}

\acmConference{PLDI'23 - EGRAPHS workshop}{June 16--21,
  2023}{Orlando, FL}
\acmPrice{}
\acmISBN{}

\begin{document}
\VerbatimFootnotes
\title{Egg-smol Python: A Pythonic Library for E-graphs}

\author{Saul Shanabrook}
\email{s.shanabrook@gmail.com}
\orcid{0000-0002-0258-4236}

\begin{abstract}
E-graphs have emerged as a versatile data structure with applications in synthesis, optimization, and verification through techniques such as equality saturation. This paper introduces Python bindings for the experimental egg-smol library, which aims to bring the benefits of e-graphs to the Python ecosystem. The bindings offer a high-level, Pythonic API providing an accessible and familiar interface for Python users. By integrating e-graph techniques with Python, we hope to enable collaboration and innovation across various domains in the scientific computing and machine learning communities. We discuss the advantages of using Python bindings for both Python and existing egg-smol users, as well as possible future directions for development.

\end{abstract}

\maketitle

\section{Introduction}
E-graphs have recently gained popularity as a versatile data structure, with applications in synthesis, optimization, and verification via equality saturation and related techniques. The growing number of open-source e-graph implementations makes it increasingly feasible to build developer-facing tools with this technology. Python has become the go-to language for scientific computing and machine learning, thanks to its extensive ecosystem of libraries and tools. Therefore, integrating e-graph techniques with the Python ecosystem can provide significant value to developers and researchers working in these fields.

Existing e-graph libraries, such as the \verb|egg| Rust library \cite{Willsey2020eggFA}, are primarily exposed through a Rust library with macros and the ability to use custom objects as expressions, as long as they implement certain interfaces for traversal. The \verb|egg-smol| library, which came after and was inspired by the \verb|egg| library, features some improvements over its predecessor\footnote{\say{egg-smol (name temporary) is a clean slate approach to this and some other shortcomings of egg. No promises on its stabilization or merging into egg yet.} Max Willsey \cite{zulip}}. One of the main differences is that it is exposed primarily through a custom s-expression Lisp-like language, and it provides concrete types instead of allowing any Rust object as expressions with traits. \verb|egg-smol| also has built-in support for typing functions and expressions.

There are a few existing e-graph libraries in Python such as Quiche \cite{quiche} and \verb|snake-egg| \cite{snake}, which is bindings for the \verb|egg| library. Many libraries in the scientific ecosystem are looking for a representation for expressing and rewriting expressions. For example, Dask \cite{matthew_rocklin-proc-scipy-2015}, a library to execute distributed computation in Python, has been looking for a way to represent high-level expression objects and execute optimizations on them, as rewrites \cite{dask-graphs}. Another library, Ibis \cite{ibis}, which presents a pandas-like Dataframe interface, but can compile to SQL, has a similar need. They are currently trying out implementing their own e-graph library, after having tried using \verb|snake-egg| \cite{ibis-egraph}. Having a standard library shared between different libraries would consolidate maintainer time and efforts, while using e-graphs would provide benefits compared to traditional rewrite systems, such as allowing less rigid rule ordering.

Integrating e-graphs in Python could also benefit end-users if the underlying e-graph library is used by multiple libraries in similar domains, allowing users or other libraries to define conversions between the constructs defined in those libraries. This would enable end-users to continue writing their business logic in the library of their choice, while using rewrite rules to convert to another representation for optimization or execution.

\section{Python Bindings for egg-smol}
This paper presents Python bindings for the egg-smol library, primarily exposing high-level constructs that mirror the s-expression library in a "Pythonic" way\footnote{The library can be installed with \verb+pip install egg-smol+, and the source available at \url{https://github.com/metadsl/egg-smol-python}. An \href{https://egg-smol-python.readthedocs.io/en/latest/tutorials/getting-started.html}{introductory tutorial} is also available on the docs.}. This high-level API is built on top of a lower-level API, exposed to Python as the \cprotect{\href{https://egg-smol-python.readthedocs.io/en/latest/reference/bindings.html}}{\verb|egg_smol.bindings| module}, which supports all the primitives of the Rust API. These bindings are created with the help of the PyO3 Rust library \cite{pyO3}.

 \begin{figure}
\begin{lstlisting}[language=Python, firstline=12, basicstyle=\ttfamily\scriptsize]
@egraph.class_
class Num(BaseExpr):
    def __init__(self, value: i64Like) -> None:
        ...

    @classmethod
    def var(cls, name: StringLike) -> Num:
        ...

    def __add__(self, other: Num) -> Num:
        ...

    def __mul__(self, other: Num) -> Num:
        ...


# expr1 = 2 * (x + 3)
expr1 = egraph.define(
    "expr1",
    Num(2) * (Num.var("x") + Num(3)),
)
# expr2 = 6 + 2 * x
expr2 = egraph.define(
    "expr2",
    Num(6) + Num(2) * Num.var("x"),
)

a, b, c = vars_("a b c", Num)
i, j = vars_("i j", i64)
egraph.register(
    rewrite(a + b).to(b + a),
    rewrite(a * (b + c)).to((a * b) + (a * c)),
    rewrite(Num(i) + Num(j)).to(Num(i + j)),
    rewrite(Num(i) * Num(j)).to(Num(i * j)),
)
egraph.run(10)
egraph.check(eq(expr1).to(expr2))
\end{lstlisting}
   \caption{High-Level Python API}
   \label{fig:high-level}
\end{figure}

A comparison between the Python API and the s-expression API can be seen in Figures \ref{fig:high-level} and \ref{fig:egg-smol}, which implements an e-graph to text the equivalence of two arithmetic  expressions.

The Python bindings utilize native Python constructs such as functions, classes, and methods, with type annotations. These constructs are translated into sort definitions and function definitions in the \verb|egg-smol| library. One notable difference between the Python bindings and the s-expression language is that variables are required to be typed in Python, whereas they are not in the s-expression language.

The requirement for typed variables in Python stems from the need for all expressions to have a known type. This aids in static analysis and code completion, as it helps disambiguate which methods are allowed based on the type. Also, in the s-expression language, everything is called as a function, which eliminates the need for typed variables. In Python, we allow classes to define the same method name, but have it translated into two different egg-smol functions. This requires knowing what class each expression has before it can be sent to \verb|egg-smol|, so we know which methods to use.

\begin{figure}
\begin{lstlisting}[language=Lisp, basicstyle=\ttfamily\scriptsize]
(datatype Math
  (Num i64)
  (Var String)
  (Add Math Math)
  (Mul Math Math))

;; expr1 = 2 * (x + 3)
(define expr1 (Mul (Num 2) (Add (Var "x") (Num 3))))
;; expr2 = 6 + 2 * x
(define expr2 (Add (Num 6) (Mul (Num 2) (Var "x"))))

(rewrite (Add a b)
         (Add b a))
(rewrite (Mul a (Add b c))
         (Add (Mul a b) (Mul a c)))
(rewrite (Add (Num a) (Num b))
         (Num (+ a b)))
(rewrite (Mul (Num a) (Num b))
         (Num (* a b)))

(run 10)
(check (= expr1 expr2))
\end{lstlisting}
   \caption{Text API}
   \label{fig:egg-smol}
\end{figure}

\section{Advantages for Python Users}
The Python bindings for the \verb|egg-smol| library provide several benefits for Python users. First, the use of built-in structures familiar to Python users makes the library more accessible and easy to use. Also, by relying on the Rust library, the Python bindings take advantage of the speed and performance that Rust offers, while allowing users to benefit from the latest innovations in the e-graph world without having to reinvent existing solutions.

Compared to the bindings for the \verb|egg| library, specifically the \verb|snake-egg| API, the Python bindings for the \verb|egg-smol| library are more opinionated. In the \verb|snake-egg| API, users could bring any Python object they wanted to map to e-graphs. In contrast, the \verb|egg-smol| bindings are more restrictive, only allowing definitions through the existing function and class wrappers.

This restriction offers several advantages, especially when considering the goal of enabling rewrites between different downstream libraries. Although these libraries may define different domains, sorts, and functions, they are represented using the same meta-structure. This uniformity makes it possible to write rewrites between them, providing a significant benefit for the community as a whole. By offering a consistent and flexible interface, the \verb|egg-smol| bindings can help create a more unified ecosystem for e-graph-based optimization and rewriting in Python.

\section{Comparison for existing egg-smol Users}
The Python interface for the \verb|egg-smol| library offers several differences to existing s-expression authors. Some aspects of the language may be more verbose in Python, such as declaring variables or adding rewrites. On the other hand, Python's operator overloading allows for more succinct mathematical expressions for custom operations.

By reusing existing Python constructs and leveraging Python's class and function structures, the Python interface provides static type checking for free, with support from existing tooling like MyPy or PyLance in Visual Studio Code. As far as static type checkers are concerned, the type of decorated functions and classes is the same as the underlying object, ensuring type checking remains consistent even though the runtime representation might differ.

\begin{figure*}
   \centering
   \includegraphics[width=1\textwidth]{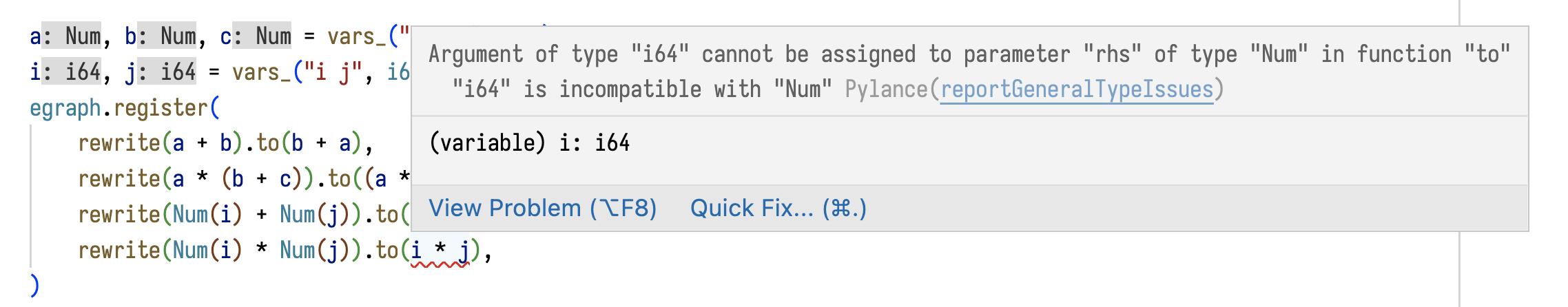}
   \caption{Type checking error in Visual Studio Code}
   \label{fig:screenshot}
\end{figure*}

Static type checking is supported not only for creating expressions but also for writing rules. Rewrites must have the same type for the left and right-hand sides, which can be verified statically by Python type checkers. See Figure \ref{fig:screenshot} for an example of the static error, if you try to replace the term \verb|Num(i) * Num(j)|, which is of type \verb|Num|, with term \verb|i * j|, which is the built-in integer type. Providing type checking on these rewrites is also the reason for the more verbose fluent syntax \verb|rewrite(lhs).to(rhs)| instead of a more succinct syntax \verb|ewrite(lhs, rhs)|. The first option allows for testing that \verb|lhs| and \verb|rhs| are the same type, whereas the second does not due to restrictions of Python's type annotations.

Auto-completion also helps users find relevant methods for rewrites, similar to how type-checking aids in code correctness. Furthermore, the Python interface allows for interactive notebook environments, such as Jupyter, to experiment with e-graph expressions and test rewrite rules on the fly.

Overall, the Python interface exposes e-graphs to a wider audience by appealing to the large user base of the Python language and providing additional guardrails for analysis and editor integration.

\section{Future Work}
The current work represents an exciting exploration of the e-graphs space with Python bindings for the \verb|egg-smol| library. However, it is yet untested in production use cases and relies on \verb|egg-smol|, which is labeled as experimental and subject to change. There are several potential directions for future work that could aid in its adoption among Python library authors:

\subsection{Support for Embedding Existing Python Types}
For Python authors dealing with existing Python objects, it would be beneficial to explore how these could be embedded in the e-graph as leaf nodes and have rules written for them. This could potentially be achieved without any modifications upstream to \verb|egg-smol| by creating a new sort for Python objects and a function to execute arbitrary Python code as a string, given some Python objects.

\subsection{Simplifying the API for User Exposure}
If the expressions and e-graphs were to be exposed to users, we would need to develop a more straightforward API for the pipeline of taking an existing expression, running some replacements on it, and outputting that back to a native Python object. This would require adding hooks to convert the output automatically to a Python object when certain methods are called.

\subsection{Prototype with an Upstream Library}
It would be helpful to find an upstream library that already uses an internal expression system with re-rewriting and prototype how this library could be used in it, as well as what improvements would be needed beforehand. Since the Ibis library is already experimenting with e-graphs, this could be a possible option.

\subsection{Exporting and Importing E-graph Descriptions}
Currently, it is not possible to write an e-graph description in Python and then use it from the s-expression language. Adding some form of export should be possible, either by emitting the s-expression language from Python or using some other machine-readable representation, such as JSON. Ideally, this could be a bi-directional transformation, allowing for the conversion between text formats and Python source.

\subsection{Relying on Python State Management and Modules}
We could rely more heavily on Python state management and modules for encapsulation. Currently, all definitions are bound to a specific e-graph instance at runtime. However, when writing a reusable library, authors might want to separate definitions between files or modules and import several of them together into an e-graph. This would make distributing and combining different modules easier.

\subsection{Interactive Visualization in Jupyter Notebooks}
By introspecting the internal state of an \verb|egg-smol| e-graph, we could create visual and interactive views of it in Jupyter notebooks. Other libraries, like Dask and Ibis, use these visualizations as educational aids to help users understand how the library processes their expressions. Developing this type of tool could allow users to step through the e-graph's execution to understand how rules are being executed and serve as an educational tool for new users to grasp the concept of e-graphs.

\section{Conclusion}
There is a growing interest in utilizing e-graphs within the Python ecosystem, and this work presents an opportunity to open up e-graphs to a wider audience. In addition to solving concrete expression optimization problems in Python data science libraries, the \verb|egg-smol| Python bindings could help make the Python data science community as a whole more resilient and connected by providing library authors with a common language to express their domains and translate between them.

Even for users who are not currently using Python, the Python bindings present an attractive authoring experience due to the extensive tooling that already exists for the Python language. By leveraging these tools and the popularity of Python, we can foster greater adoption of e-graph techniques and facilitate collaboration and innovation across various domains.

\begin{acks}
Thank you to the Recurse Center for providing the community and support to do this work. In particular, thank you Sean Aubin for your feedback on this proposal, which was invaluable in the editing process. None of this would be possible without all of the work on the underlying Rust libraries, by Max Willsey and others. Thank you for fielding my questions about \verb|egg-smol| and be open to collaboration on it, even at such a young stage.

\end{acks}

\bibliographystyle{ACM-Reference-Format}
\bibliography{sample-base}


\end{document}